\documentstyle[aps,twocolumn,epsf]{revtex}

\begin{document} 

\title{Magnetization switching in a Heisenberg model for small
  ferromagnetic particles}
\author{D.~Hinzke and U.~Nowak}
\address{Theoretische Tieftemperaturphysik,
  Gerhard-Mercator-Universit\"{a}t-Duisburg, 47048 Duisburg/ Germany\\ 
  e-mail: uli@thp.uni-duisburg.de }

\date{\today}
\maketitle

\begin{abstract}
  We investigate the thermally activated magnetization switching of
  small ferromagnetic particles driven by an external magnetic field.
  For low uniaxial anisotropy the spins can be expected to rotate
  coherently, while for sufficient large anisotropy they should behave
  Ising-like, i.~e.~, the switching should then be due to nucleation.
  We study this crossover from coherent rotation to nucleation for the
  classical three-dimensional Heisenberg model with a finite
  anisotropy.  The crossover is influenced by the size of the
  particle, the strength of the driving magnetic field, and the
  anisotropy.  We discuss the relevant energy barriers which have to
  be overcome during the switching, and find theoretical arguments
  which yield the energetically favorable reversal mechanisms for
  given values of the quantities above. The results are confirmed by
  Monte Carlo simulations of Heisenberg and Ising models.
\end{abstract}

Pacs: 75.10.Hk, 75.40.Mg, 64.60.Qb


\section{Introduction}
The size of magnetic particles plays a crucial role for the density of
information storage in magnetic recording media. Sufficiently small
particles become single-domain particles, which improves their quality for
magnetic recording. On the other hand, when the particles are too
small they become superparamagnetic and no information can be stored
(see e.~g.~\cite{chantrell} for a review). Hence, much effort has
recently been focused on the understanding of small magnetic particles,
especially since recent experimental techniques allow for the
investigation of isolated single-domain particles
\cite{salling,lederman,wernsdorfer,wernsdorfer2}.

In this paper focus is on the reversal of ferromagnetic
particles of finite size.  We investigate the influence of the size
and anisotropy of the particle on the possible reversal mechanisms,
two extreme cases of which are coherent rotation and nucleation.
The latter mechanism has been a subject of common interest in recent
years \cite{stauffer,tomita,rikvold,acharyya,garcia,richards}, studied
mainly theoretically in Ising models, which can be interpreted as a classical
Heisenberg model in the limit of infinite anisotropy. It is the aim of
this paper to study the crossover from magnetization reversal due to
nucleation for high anisotropy to coherent rotation \cite{neel,brown}
for lower anisotropy.

Throughout the paper we will consider a finite, spherical three-
dimensional system of magnetic moments.  These magnetic moments may
represent atomic spins or block spins following from a coarse
graining of the physical lattice \cite{nowak}.  Our system is defined
by a classical Heisenberg Hamiltonian,
\begin{equation}
  {\cal H} = - J \sum_{\langle ij \rangle} {\bf S}_i \cdot {\bf S}_j
  -d \sum_i (S^z_i)^2 -{\bf B} \cdot \sum_i {\bf S}_i.
\label{e:ham}
\end{equation}
where the ${\bf S}_i$ are three-dimensional vectors of unit length.
The first sum which represents the exchange of the spins, is over
nearest neighbors with the exchange coupling constant $J$. The second
sum represents an uniaxial anisotropy which favors the $z$-axis as
easy axis of the system (anisotropy constant $d>0$). The last sum is
the coupling of the spins to an applied magnetic field, where ${\bf
  B}$ is the strength of the field times the absolute value of the
magnetic moment of the spin. We neglect dipolar interaction. Although
in principle it is possible to consider dipolar interaction in a Monte Carlo
simulation \cite{nowak} this needs much more computational effort due
to the long range of the dipolar interaction and, hence, exceeds
current computer capacities . Therefore, the validity of our results
is restricted to particles which are small enough to be single-domain
in the remanent state \cite{aharony}.

In the following, we will investigate the thermally activated reversal
of a particle which is destabilized by an magnetic field pointing in a
direction antiparallel to the initial magnetization which is
parallel to the easy axis of the system. Due to the finite temperature
and the magnetic field, after some time the particle will reverse its
magnetization, i.~e. the $z$-component of the magnetization will
change its sign.

In section \ref{s:theory}, we determine the energy barriers which have
to be overcome by thermal fluctuations for the two cases of coherent
rotation and nucleation within a classical theory.  By a comparison of
the energy barriers, we derive where the crossover from one
mechanism to the other occurs.

In section \ref{s:numerics} we compare our theoretical considerations
with numerical results from Monte Carlo simulations of Heisenberg and
Ising models, and we relate the lifetime of the metastable state to
the theoretical energy barriers for the different reversal mechanisms.

\section{Theory}
\label{s:theory}
\subsection{Coherent rotation}
We give here a brief summary of the results of the theory of N\'{e}el
\cite{neel} and Brown \cite{brown}, since we need this concept for the
further progress of our theoretical considerations.

Let us consider a spherical, homogeneously magnetized particle of
radius $R$. The simplest theoretical description for the reversal of
such a particle is to assume that the reversal mechanism is coherent
rotation, i.~e. a uniform rotation of all spins of the particle. This
reversal process can be described by an angle of rotation $\theta$
between the easy axis of the system -- which in our case will be
antiparallel to the direction of the magnetic field -- and the
magnetization of the particle. The increase of the energy during the
reversal is then
\begin{equation}
  \Delta E = -d V \cos^{2}{\theta} - BMV \cos{\theta}.
\end{equation}
Since this equation should be comparable to Eq.~\ref{e:ham}, the
anisotropy constant $d$ is an anisotropy energy per unit cell (spin),
and $V= (4/3) \pi R^3$ is the volume of the particle as a number of
unit cells. $B$ is - as before - the absolute value of the applied
field times the magnetic moment of a unit cell and, hence, $M$ the
spontaneous magnetization per magnetic moment. The energy barrier
which has to be overcome is due to the anisotropy of the system. It is
the maximum of $\Delta E$ with respect to $\theta$:
\begin{equation}
  \Delta E_{cr}  =  \frac{4 \pi R^3d}{3}-{\frac {4\pi R^3BM}{3}}
  +\frac{\pi R^3B^2M^2}{3d}
  \label{e:ecr}
\end{equation}
The corresponding lifetime of the metastable state is then
\begin{equation}
  \tau \sim \exp \left( \frac{\Delta E_{cr}}{T} \right)
  \label{e:activation}
\end{equation}
for temperature $T \ll \Delta E_{cr}$.  The two equations above are
physically relevant only for
\begin{equation}
  d > M B/2.
  \label{e:thermal}
\end{equation}
Otherwise there is no (positive) energy barrier, and hence the reversal
is spontaneous without the need for thermal activation. This is the
region of nonthermal reversal.

\subsection{Nucleation}
\begin{figure}
  \begin{center}
    \hspace*{1.cm}
    \epsfxsize=5cm
    \epsffile{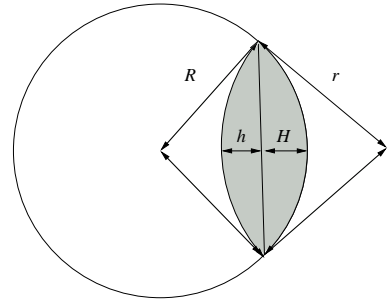}
  \end{center}
  \caption{The nucleation of a droplet at the boundary of a spherical
    particle.}
  \label{f:droplet}
\end{figure}

For a system with a sufficient large anisotropy, it might be
energetically favorable to divide into parts with opposite directions
of magnetization parallel to the easy axis in order to minimize the
anisotropy energy barrier. This kind of reversal mechanism is called
nucleation \cite{becker} (see \cite{rikvold} for a recent review). The
simplest case of a reversal process driven by nucleation for a system
of finite size is the growth of one single droplet starting at one
point of the boundary of the system (see also \cite{richards} for a
corresponding calculation for a two-dimensional system). Due to the
growth of the droplet a domain wall will cross the system, and the
energy barrier which has to be overcome is caused by the domain-wall
energy. We assume that the domain wall will have a curvature defined
by a radius $r$ (see Fig.~\ref{f:droplet}).

Then the surface of the domain wall is $F = 2 \pi rh$ and the volume
of the droplet (the shaded region in Fig.~\ref{f:droplet}) is
$V_d = \pi H^2 (3R-H)/3 + \pi h^2 (3r-h)/3$.
The energy increase during the reversal of the particle is
\begin{equation}
  \Delta E = 2 \sigma F - 2 M B V_d,
\end{equation}
where $\sigma$ is the energy density of the domain wall. Furthermore,
it is $h = r - \sqrt {r^2 -2HR + H^2}$, since the quantities in
Fig.\ref{f:droplet} are not independent. Hence the energy increase
can be expressed in terms of $H$, which is a measure for the
penetration depth of the domain wall and its curvature $r$. These two
quantities define the geometry of the droplet.  Next we determine that
curvature $r_{\mbox{min}}$ which minimizes the energy increase by the
condition $\frac {\partial {\Delta E}} {\partial{r}} = 0$, yielding the
physically relevant solution
\begin{equation}
  r_{\mbox{min}} = \frac {2\sigma} {MB}
  \label{e:rmin}
\end{equation}
which is also the radius of a critical droplet for classical nucleation in a
bulk material.

The energy barrier $\Delta E_w$ which has to be overcome during the
reversal is the maximum of the energy increase with respect to $H$.
From the condition $\frac {\partial {\Delta E}} {\partial{H}} = 0$,
the physically relevant solution
\begin{equation}
  H_{\mbox{max}} = R \Big(1 - \frac{x}{\sqrt{x^2+4}} \Big)
\end{equation}
follows with $x = MBR/\sigma$. 
Inserting the two conditions above into the formula for the energy
increase yields the energy barrier for nucleation
\begin{eqnarray}
  \Delta E_n & = &  \frac{4\pi R^2 \sigma}{3x^2\sqrt{x^2+4}} \Big(x^4
   + (4-x^3)\sqrt{x^2+4} +2x^2 -8 \Big) \nonumber \\
   && 
   \label{e:edw}
\end{eqnarray}
This expression has two important limits. The first is the limit of
infinite system size,
\begin{equation}
  \lim_{R \rightarrow \infty} \Delta E_n = \frac{16 \pi \sigma^3}{3M^2B^2},
\end{equation}
where we obtain half of the energy barrier of the classical nucleation
theory for a bulk system. The reduction by a factor of $\frac{1}{2}$ is due
to the fact that for open boundary conditions only one-half of a
critical droplet has to enter the system from the boundary.

The other interesting limit is that of small magnetic fields, where
Eq.\ref{e:edw} can be expanded with respect to $x = MBR/\sigma$ resulting in 
\begin{equation}
  \Delta E_n \approx 2 \pi R^2 \sigma - \frac{4 \pi B R^3 M}3 
  +\frac {3\pi B^2R^4 M^2}{8\sigma} + \ldots
\end{equation}
This means that for a small field $B$ the energy barrier of a nucleation
process is the energy of a flat domain wall in the center of the
particle plus corrections which start linearly in $B$.  In contrast to
the N\'{e}el-Brown theory, here, for vanishing magnetic field, the
energy barrier is proportional to the cross-sectional area of the
particle rather than its volume (see also the work of Braun
\cite{braun}), which consequently reduces the coercivity of the particle.

\subsection{Comparison of coherent rotation and nucleation}
\begin{figure}
  \begin{center}
    \epsfysize=4.5cm
    \epsffile{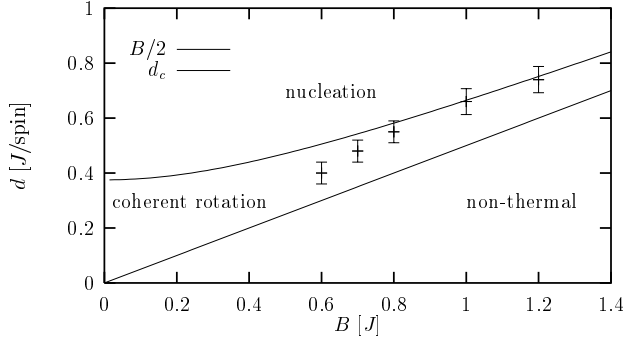}
  \end{center}
  \caption{Diagram showing the regions of different reversal mechanisms for a
    particle of size $R=4$ spins, with $\sigma = J/$spin and $M=1/$spin.
    The points are results from Monte Carlo simulations (see Section
    \ref{s:numerics}).}
  \label{f:diagram}
\end{figure}

Comparing the two energy barriers for coherent rotation
(Eq.\ref{e:ecr}) and nucleation (Eq.\ref{e:edw}), we can evaluate which
reversal mechanism has the lower activation energy for a given set of
values of $R$, $B$, $\sigma$, and $d$. A corresponding diagram is shown
in Fig.~\ref{f:diagram} for a system of radius $R=4$ spins, where we
set $M=1/$spin and $\sigma = J/\mbox{spin}$.
For large anisotropy the reversal is dominated by nucleation -- the
particle behaves like an Ising system.  The crossover line
$d_c(B,\sigma,R)$ which separates the region of reversal by nucleation
from the region of coherent rotation can be determined by the
condition that here the energy barriers for a nucleation process and
for coherent rotation are equal.  This condition results in
\begin{eqnarray}
\label{e:dw}
  d_c & = & \frac{\sigma}{R} \Big(\frac{2}{x^2}
  +\frac{1}{\sqrt{x^2+4}} \big(1 -\frac{4}{x^2} +\frac{x^2}{2}\big) \nonumber\\
        && + \frac{1}{x^2} \sqrt{-3x^2+8 + (2x^2-4)\sqrt{x^2+4}} \Big).
\end{eqnarray}
For vanishing magnetic field, the formula above has a finite limit, 
\begin{equation}
  \lim_{B\rightarrow 0} (d_c) = \frac{3\sigma}{2R}.
  \label{e:dclb}
\end{equation}
For large particle size, Eq.~\ref{e:dw} has the simple asymptotic form
\begin{equation}
  \lim_{R\rightarrow \infty} (d_c) = \frac{BM}{2}
\label{e:nt}
\end{equation}
which, interestingly, is also the limit for 
nonthermal reversal (see Eq.~\ref{e:thermal}). This means that
for increasing particle size the region where a thermally activated
reversal by coherent rotation occurs vanishes. For an infinite particle
size the reversal is always either nonthermal or it is driven by
nucleation -- depending only on the ratio of the magnetic field
to the anisotropy.

All the considerations above can be expected to be relevant only for
sufficiently low temperatures. For higher temperatures the situation
is more complicated. That is, for the nucleation regime, a crossover from
single-droplet to multi-droplet nucleation -- with different energy
barriers) -- is discussed in the literature (see
\cite{binder,ray,sides} and references therein).

Apart from that, in a system with a given finite anisotropy the domain
walls may be extended to a certain domain wall width $\xi$. For large
anisotropy $\xi$ becomes as small as one lattice constant. This is the
Ising case, where the domain-wall energy density is $\sigma \approx
J/$spin. For smaller anisotropy the domain walls become more extended,
and $\sigma $ decreases. Hence the crossover from nucleation to
coherent rotation may be softened by the occurrence of extended domain
walls. In this sense, pure coherent rotation could also be
interpreted as a domain-wall-driven reversal, where the width of the
domain wall is larger than the particle size.  Obviously, our
theoretical considerations discuss only two extreme cases. How
realistic they are has to be tested numerically.

\section{Monte Carlo Simulation}
\label{s:numerics}
\subsection{Method}
Due to the many degrees of freedom of a spin system, numerical methods
have to be used for a detailed microscopic description of the system.
Since we are especially interested in the thermal properties of the
system, we use Monte Carlo methods \cite{heermann} for the simulation
of the magnetic particle. Although a direct mapping of the time scale
of a Monte Carlo simulation on experimental time scales is not
possible, this method provides information on the dynamical behavior of
the system since it solves the master equation for the irreversible
behavior of the system \cite{reif}.

We consider spins on a simple cubic lattice of size $L\times L \times
L$, and simulate spherical particles with radius $R = L/2$ and open
boundary conditions on this lattice.

One single spin flip of our Monte Carlo procedure consists of three
parts.  First, a spin is chosen randomly and a trial step is made (the
role of which we will discuss below). Second, the change of the energy
of the system is computed according Eq.~\ref{e:ham}.  Third, the trial
step is accepted with the probability from the heat-bath algorithm.
Let us call one sweep through the lattice and performing the procedure
explained above once per spin one Monte Carlo step (MCS).

Since we are interested in different reversal mechanisms, we designed a
special algorithm which can simulate all of them efficiently.  We use
three different kinds of trial steps: First, a trial step in any spin
direction uniformly distributed in spin space.  This step does not
depend on the initial direction of the spin. It samples the whole
phase space efficiently and guarantees ergodicity.  Second, a small
step within a limited circular region around the initial spin
direction. This step can efficiently simulate the coherent rotation.
Third, a reflection of the spin. This step guarantees that, in the
limit of large anisotropy, our algorithm crosses over to an efficient
simulation of an Ising-like system.  For each Monte Carlo step we use
one of these different trial steps.  Our algorithm then consists of a
series of Monte Carlo steps using the different trial steps above.
Altogether, our algorithm is ergodic, and it guarantees that all
possible reversal mechanisms may occur in the system and can be
simulated efficiently. As we tested by comparing simulations with 
different combinations of trial steps, and as we will demonstrate in 
the section \ref{s:results_heise}, for a two-spins system, this
algorithm does not artificially change our results.

Simulations of Heisenberg systems are much more time consuming, than
e.~g.~, those of Ising systems, since the Heisenberg system has many more
degrees of freedom. Apart from that, to obtain results which are
comparable to our theoretical considerations we have to perform
simulations in the limit of low temperatures, $T \ll T_c$, where the
critical temperature $T_c$ is 1.44$J$ for anisotropy $d=0$ \cite{ciftan}.
Here, the metastable lifetimes are long -- for single runs up to
$5\times10^7$MCS in our simulation.  Therefore, for the Heisenberg
system we had to restrict ourselves to rather small system sizes,
$L=4,8,12$.  However, we tried to minimize the statistical error by
performing an average over many Monte Carlo runs ($100 \ldots 1000$).
Since the theoretical considerations which we want to proof are for
finite system sizes, and since, hence, the radius of the particle is a
variable of the theory we belive that the rather small system sizes of
the simulation are no disadvantage.  Apart from that, for comparison
we also performed Monte Carlo simulations of an Ising model where we
also used larger system sizes of up to $L=28$.

The simulations were performed on an IBM-RS6000 workstation cluster and
on two Parsytec CC parallel computers (8 and 24 PPC604 nodes,
respectively).

\subsection{Results for the Heisenberg system}
\label{s:results_heise}
We start our simulation with an initial spin configuration where all
spins are pointing up (all spins ${\bf S}_i = (0,0,1)$). The magnetic
field ${\bf B} = (0,0,-B)$ destabilizes the system, and after some time
the magnetization of the system will reverse. The metastable lifetime
$\tau$ is defined by the condition $M_z(\tau) = 0$ where $M_z$ is the
$z$-component of the magnetization ${\bf M} = (1/N) \sum_i {\bf S}_i$.

\begin{figure}
  \begin{center}
    \epsfysize=4.5cm
    \epsffile{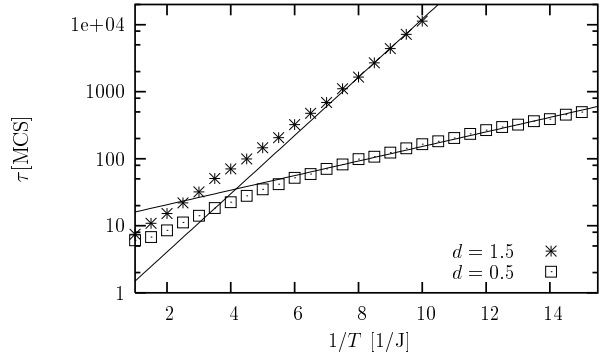}
  \end{center}

  \vspace{-5mm}
  \caption{Metastable lifetime $\tau$ vs. $1/T$ for a system of two
    spins. $B = 0.5J$. Two different anisotropies, $d = 1.5$/spin
    (nucleation) and $d = 0.5J$/spin (coherent rotation). The solid
    lines correspond to Eq.~\ref{e:activation}, with the energy
    barriers explained in the text.}
  \label{f:two_spins}
\end{figure}

First, we tested our algorithm by simulating the simplest
imaginable system that can show both coherent rotation and nucleation
-- a two-spin system. Here the energy barrier is $\Delta E_n =
2(J-B)$ for nucleation (i.~e.~, the spins are antiparallel during the
reversal) and $\Delta E_{cr} = 2d(1-B/(2d))^2$ for coherent rotation
(i.~e.~, the spins are always parallel during the reversal). For low
temperatures we expect a behavior following thermal activation as in
Eq.~\ref{e:activation} with the energy barriers above.
Figure~\ref{f:two_spins} demonstrates that in the limit of low
temperatures, we actually obtain constant slopes for the $\ln \tau$ vs.
$1/T$ data, and the slopes agree perfectly with the theoretical energy
barriers above. Hence our simulation is in agreement with the
theoretical expectations.

Now we turn to the simulation of larger systems.
\begin{figure}
  \begin{center}
    \hspace{2mm}
    \epsfysize=4.cm
    \epsffile{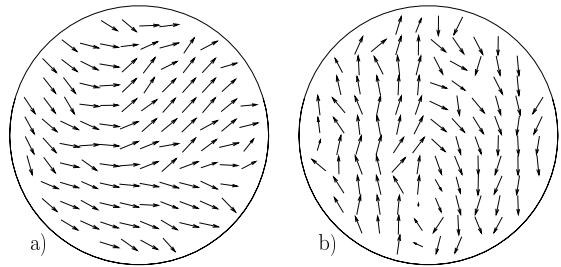}
  \end{center}
  \vspace{-2mm}
  \caption{Snapshots of simulated spin configurations at the lifetime
    $\tau$. Shown is one central plane of systems of size
    $R=6$ spins. $B = 0.7J$.  a) Coherent rotation ($d = 0.35J$/spin,
    $T=0.09J$). b) Nucleation ($d = 0.7J$/spin, $T = 0.45J$).}
  \label{f:spin}
\end{figure}

Fig.~\ref{f:spin} shows spin configurations of simulated systems of
size $R=6$ spins at the metastable lifetime $\tau$. For simplicity,
only one central plane of the three-dimensional system is shown. The
$z$-axis of the spin components is pointing up.  For sufficient low
anisotropy (Fig.\ref{f:spin}a) the spins rotate nearly coherently. At
the metastable lifetime $\tau$ the magnetization vector of the system
points to any given direction in the $x-y$-plane.  Therefore, as
horizontal component of the spins we show here that component of the
$x-y$-plane of the spin space that has the largest
\begin{figure}
  \epsfxsize=8.cm
  \epsffile{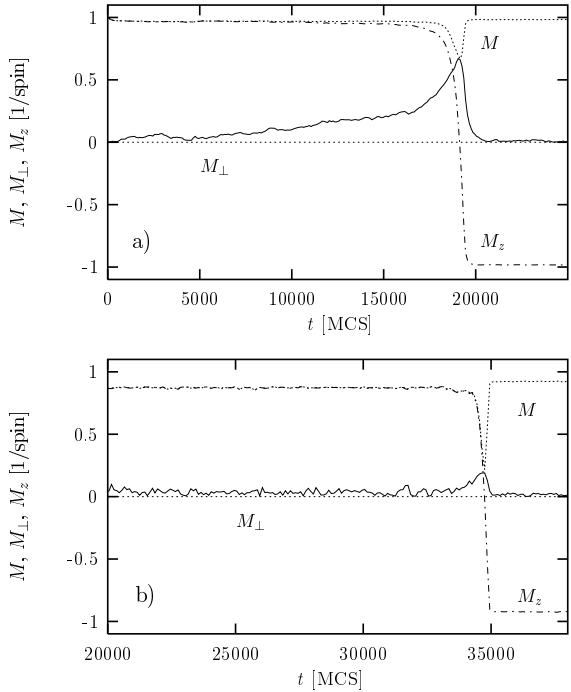}

  \vspace{2mm}
  \caption{$M$, $M_{\perp}$, and $M_z$ of one $R=6$ spins - system
    as in Fig.\ref{f:spin}: $B=0.7J$.  a) Coherent rotation ($d =
    0.35J$/spin, $T=0.09J$). b) Nucleation ($d = 0.7J$/spin), $T =
    0.45J$.}
  \label{f:mag_curves}
\end{figure}
contribution. For larger anisotropy (Fig.\ref{f:spin}b) the reversal
is driven by nucleation. Since it is $M_z(\tau)=0$, the domain wall at
that time is in the center of the system dividing the particle into
two oppositely magnetized parts of equal size.

Fig.~\ref{f:mag_curves} shows the corresponding time dependence of the
$z$-component of the magnetization, its absolute value $M = |{\bf M}|$,
and its planar component $M_{\perp} = \sqrt{M_x^2+M_y^2}$ for the same
simulation from which the spin configurations of Fig.\ref{f:spin} stem.  For
the case of coherent rotation there is a continuous growth of the
planar component of the magnetization during the reversal, while the
absolute value of the magnetization remains nearly constant -- apart
from a small dip at the lifetime $\tau \approx 19000$MCS. For the case
of nucleation the planar component of the magnetization is nearly
constant zero except of a small hump at the lifetime $\tau \approx
34000$MCS. Here the absolute value of the magnetization breaks down.

These results lead us to the following approach to characterize the
reversal mechanisms numerically: we determine the absolute value of
the magnetization at the lifetime, $M(\tau)$. In order to obtain
reasonable results we have to take an average over many runs, so that
we define a quantity $\mu = [M(\tau)]$ where the square brackets
denote an average over many Monte Carlo runs (or systems). This
quantity should go to zero for a nucleation-driven reversal, and should
be finite for coherent rotation. The maximum value of $\mu$ in the
limit of low anisotropy should be the spontaneous magnetization.

\begin{figure}
  \epsfxsize=8.0cm
  \epsffile{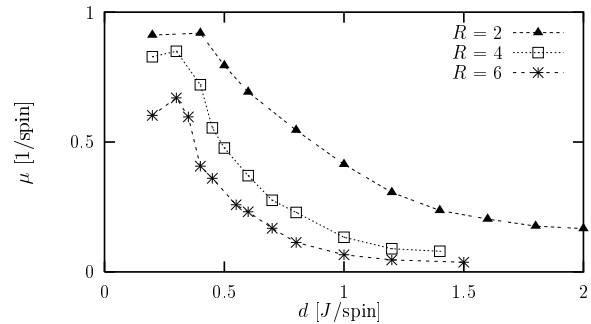}
  
  \vspace{2mm}
  \caption{$\mu$ vs.~anisotropy for different system sizes. $B = 0.7J$,
    $T = 0.71 \ldots 0.04J$ depending on $d$. }
  \label{f:total_mag}
\end{figure}

\begin{figure} \vspace{5mm}
  \epsfxsize=7.0cm
  \epsffile{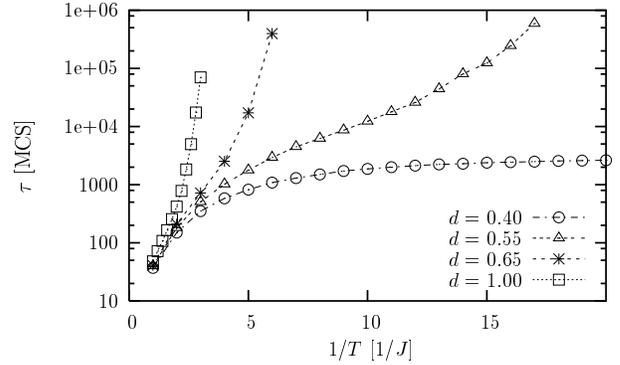}
  
  \vspace{2mm}
  \caption{Metastable lifetime $\tau$ vs. $1/T$ for four
    different anisotropies. System size $R=4$ spins and $B = 1J$.}
  \label{f:tau_heise}
\end{figure}

In order to confirm our theoretical results numerically, we simulated
$\mu$ for different values of the anisotropy $d=0.2 \ldots 2J/$spin, the
magnetic field $B = 0.7 \ldots 1.2J$, and the system size $R=2 \ldots
6$ spins. We took an average over 100 ($R=6$) to 1000 ($R=2$) runs.
Fig.~\ref{f:total_mag} shows the results for the anisotropy dependence
of $\mu$ for different system sizes and the lowest temperatures that
have been simulated ($T = 0.71 \ldots 0.04J$ depending on $d$). The
influence of the temperature on the simulations will be discussed
later in connection with Fig.~\ref{f:tau_heise}.

As expected, for small anisotropy, $\mu$ tends to a finite limit while
with increasing anisotropy the curves converge to zero.  This effect
is stronger the larger the system size is --- a behavior that
appears to be analogous to the finite size behavior of a system
undergoing a phase transition. The corresponding finite-size scaling
ansatz \cite{heermann} is
\begin{equation}
  \mu = R^{-\beta/\nu} \tilde{\mu}^{\pm}\left(|d_{\infty} - d| R^{1/\nu}\right)
\end{equation}
where the scaling function $\tilde{\mu}^{\pm} \sim x^{\beta}$ for $x
\rightarrow \infty$ so that 
\begin{figure}
  \vspace{3mm}
  \epsfxsize=8.0cm
  \epsffile{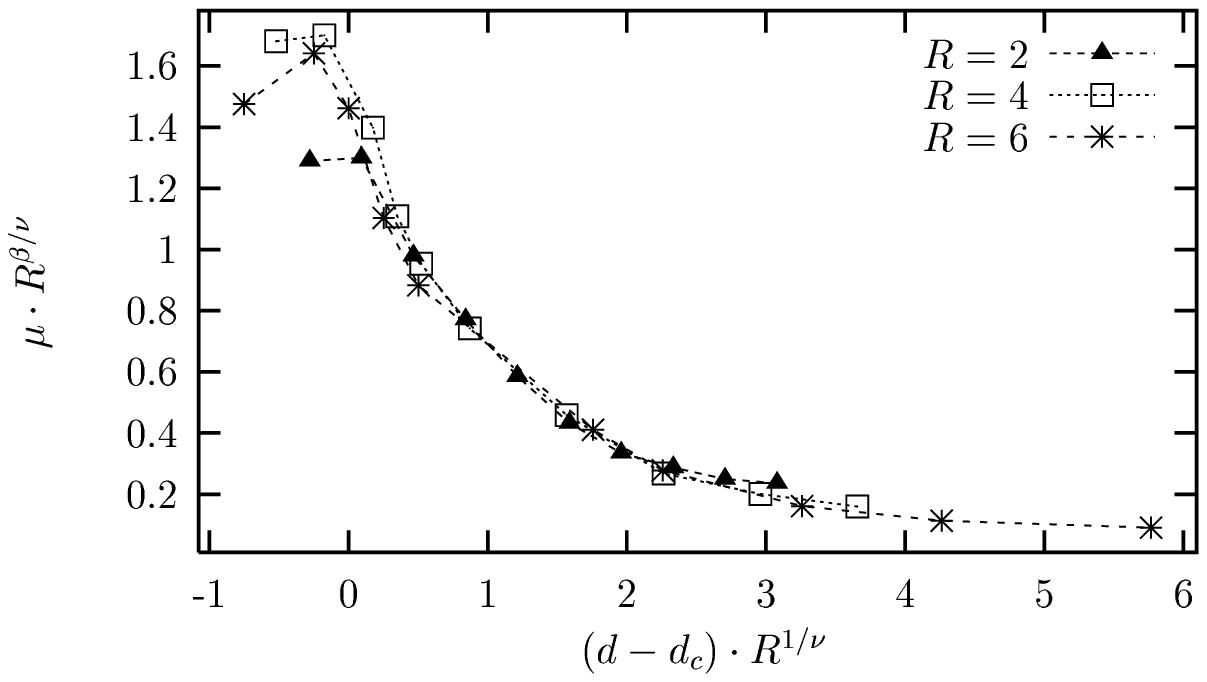}
  
  \vspace{2mm}
  \caption{Scaling plot of $\mu$ vs. anisotropy for different system
    sizes. Data correspond to Fig.~\ref{f:total_mag}}
  \label{f:scaling}
\end{figure}
in the limit of infinite system size it is $\mu \sim (d_{\infty}
-d)^{\beta}$ for $d < d_{\infty}$. In order to test if this scaling
form can be applied to the data shown in Fig.~\ref{f:total_mag}, we
present a corresponding scaling plot in Fig.~\ref{f:scaling}. The data
collapse rather well, using $d_{\infty} = B/2 = 0.35J/\mbox{spin}$,
$\beta/\nu = 0.5 \pm 0.1$, and $1/\nu = 0.9 \pm 0.1$.

Obviously, in the limit $R \rightarrow \infty$ $\mu(d)$ behaves like
an order parameter at a second-order phase transition: it is zero for
$d > d_{\infty}$ and finite for $d < d_{\infty}$, following $\mu \sim
(d_{\infty}-d)^{\beta}$.

The fact that for infinite system size the transition occurs at
$d_{\infty} = B/2$ is in agreement with our theoretical considerations,
since for $R \rightarrow \infty$ the region of thermally activated
coherent rotation vanishes and the crossover from nucleation to
(nonthermal) coherent rotation occurs at $d_c(R \rightarrow \infty) =
B/2$ (Eq.~\ref{e:nt}). That is, only for $d<B/2$ will the particle
rotate coherently, and $\mu$ must be finite.  However, to what degree the
crossover from nucleation to coherent rotation for infinite system
size may be described as a phase transition must be left for future
research.

In the following we will restrict ourselves on systems of finite size.
In order to differentiate numerically between the two reversal
mechanisms for systems of finite size, we use a criterion that also comes
from a study of phase transitions: we define the inflection point of
the curves $d(B)$ as that value $d_c(B,R)$ where the crossover from
nucleation to coherent rotation occurs. These points are shown in
Fig.~\ref{f:diagram}. For large $B$ they agree very well with the
theoretical line. For lower $B$ the numerical values are slightly too
small. This systematic deviation might be due to the fact that the
theoretical energy barrier for nucleation is overestimated assuming
$\sigma = J/$spin: due to occurrence of extended domain walls for
lower anisotropy the domain wall energy might be reduced. This also reduces
the energy barrier of the nucleation process and consequently,
here crossover to nucleation occurs earlier.

Apart from the numerical determination of the cross overline $d_c$
discussed above, we also tried to compute the relevant energy barriers
directly.  During the simulation, temperature plays a crucial role.
The larger $d$ is, the larger is the energy barrier which has to be
overcome by thermal activation and, hence, the higher the
temperature has to be during the simulation in order to obtain results within a
given computing time. On the other hand, we have to simulate as low
temperatures as possible to see the behavior that is described by our
theoretical considerations.  Therefore, varying the anisotropy, we have
to adjust the temperature.  Fig.~\ref{f:tau_heise} shows the
temperature dependence of the metastable lifetime for different
anisotropies. To save computer time, instead of an average of the
lifetimes of the individual Monte Carlo runs, we calculated the median
\cite{stauffer}. In the limit of low temperature we expect a behavior
following thermal activation as in Eq.~\ref{e:activation}, with the
energy barrier following from the theoretical consideration in section
\ref{s:theory}.

\begin{figure}
  \epsfxsize=8.0cm
  \epsffile{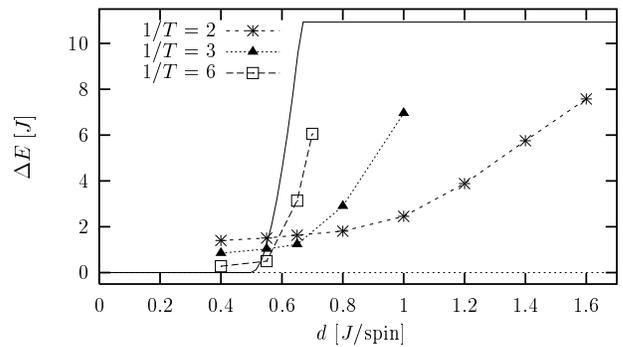}
  
  \vspace{2mm}
  \caption{Comparison of the theoretical curve for the $d$-dependence
    of the energy barrier with the numerical data for different
    temperatures following from Fig.~\ref{f:tau_heise}.}
  \label{f:energy_d}
\end{figure}

As Fig.~\ref{f:tau_heise} shows, this dependence (i.~e.~, constant
slopes in Fig.~\ref{f:tau_heise} for low $T$) can hardly be observed.
All curves have a finite curvature even for the lowest simulated
temperature except of that for $d = 0.4J$/spin, where the energy barrier is
zero. We conclude that we could not reach low enough temperatures
within the simulations and, hence, we analyse our data in the
following way: We take the local slope of $(\ln \tau)(1/T)$ as
temperature-dependent energy barrier. These energy barriers versus $d$
are shown in Fig.~\ref{f:energy_d} for three different temperatures,
and they are compared with the theoretical results.
The magnetic field is $B=1J$.  Hence, for $d < 0.5J/$spin -- the
nonthermal region (Eq.~\ref{e:nt}) -- it is $\Delta E = 0$. In the
regime for thermally activated coherent rotation, i.~e.~, between $d <
0.5J/$spin and the crossover anisotropy $d_c$ (Eq.~\ref{e:dw}), the
energy barrier increases following Eq.~\ref{e:ecr}. Above $d_c$ which
is roughly $0.65J/$spin, here the energy barrier remains constant since
in the nucleation regime it does not depend on $d$ (Eq.~\ref{e:edw}).

Comparing our numerical results with this theoretical curve, we find
agreement as far as the crossover from nonthermal reversal to
thermally driven coherent rotation is concerned. Also, we observe that
in principal the numerical results seem to converge with decreasing
temperature. However, within these simulation we cannot confirm the
theoretical curve very accurately, especially the asymptotic behavior
of the nucleation regime. We conclude that the main reason for this
quantitative deviation is that we do not obtain the asymptotic
low-temperature energy barriers, since these depend still on the temperature.

One additional reason for deviations in the coherent rotation regime
might be that, in order to compare the results of our simulation with
the theoretical results, we had to estimate the domain-wall energy density
$\sigma$, which here we simply set to $\sigma = J/$spin. This estimate
might be to large for a Heisenberg system which can develop extended
domain walls with a lower domain-wall energy. We could try to fit
$\sigma$ in such a way that it we obtain a reasonable agreement with the
numerical data, but this would not solve the problem mentioned above,
namely, that we are not in the asymptotic low-temperature regime. 

\subsection{Results for the Ising system}
In order to confirm our theoretical results for the energy barrier of
the nucleation regime, it is much more straightforward to 
simulate an Ising system directly instead of a Heisenberg system with large
anisotropy. Therefore, we performed a standard Monte Carlo simulation
of an Ising system defined by the Hamiltonian

\begin{equation}
  {\cal H} = - J \sum_{\langle ij \rangle} S_i S_j - B \sum_i S_i
\end{equation}

with $S_i = \pm 1$. As before, we simulated spherical particles with
radius $r = L/2$ on a simple cubic lattice of size $L\times L \times
L$ with open boundary conditions. For the Ising system, $L$ was varied
from $L = 4$ to $L = 32$. We used the same methods as above, performing
averages over 50-100 Monte Carlo runs depending on the system size.

\begin{figure}
  \epsfxsize=7.5cm
  \epsffile{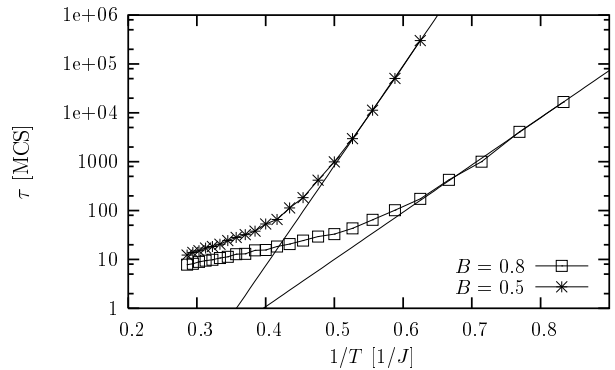}
  
  \vspace{2mm}
  \caption{Metastable lifetime $\tau$ vs. $1/T$ for two different 
   magnetic fields. System size $R=4$ spins.}
  \label{f:tau_ising}
\end{figure}

Fig.~\ref{f:tau_ising} shows the resulting temperature dependence of
the metastable lifetime. In our data for the Ising system, a thermal
activation corresponding to Eq.\ref{e:activation} can be much better 
extracted than for the Heisenberg system. For low enough temperatures,
straight lines can be fitted to the data the slope of which determine the
activation barrier $\Delta E_n$.

For a comparison of our numerically determined activation barriers with
Eq.\ref{e:edw} for the theoretical energy barriers of the nucleation
process, once more we have to estimate the domain-wall energy density
$\sigma$. In an Ising system, the domain-wall width is reduced to one
lattice constant and, hence, the domain-wall energy cannot be reduced
by an extended width of the wall. But, on the other hand, on a cubic
lattice the energy of a domain wall per spin depends on the direction
of the domain wall with respect to the axis of the lattice. This is
$J/$spin for a wall parallel to the axis but, larger for walls in
diagonal directions. Thus, we can expect $\sigma$ to be little larger
than $J/$spin, and in the following it is set to $\sigma =
1.2J$.

\begin{figure}
  \epsfxsize=8.4cm
  \epsffile{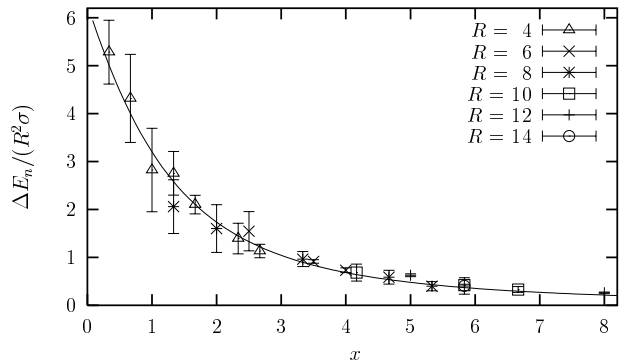}
  
  \vspace{2mm}
  \caption{Comparison of the theoretical curve for the energy barrier
    (Eq.~\ref{e:edw}) with the numerical data for different system
    sizes and $B = 0.1 \ldots 0.8$. $x = MBR/\sigma$.}
  \label{f:energy_x}
\end{figure}

Fig.~\ref{f:energy_x} compares our numerical data with
Eq.~\ref{e:edw}.  Since $\Delta E_n/(R^2\sigma)$ depends only on the
variable $x = MBR/\sigma$, the data for different system sizes collapse
on one single curve, and the agreement of our numerical data with the
theoretical curve is satisfactory.

\section{Conclusions}
We investigate the magnetization reversal of a classical Heisenberg
system for single-domain ferromagnetic particles. Varying the
anisotropy one expects different reversal mechanisms, the extreme cases
of which are coherent rotation and nucleation, which in the case of a
single-droplet-nucleation is a reversal by domain-wall motion. We
studied the crossover from switching due to nucleation for high
anisotropy, high fields, and large systems to coherent rotation for
lower anisotropy, lower fields, and smaller systems. By a comparison of
the relevant energy barriers, we derive a formula which estimates where
the crossover from the one mechanism to the other occurs.

If we insert the material parameters for CoPt \cite{nowak}, $\sigma =
0.004\mbox{J/m}^2$, and $d = 200$kJ/$\mbox{m}^3$ in Eq.~\ref{e:dclb},
we find that for vanishing magnetic field a reversal by nucleation
should occur for particles with a radius larger than 15nm. While at
the moment we are not aware of any direct measurement of this
crossover, experimental hints of the occurrence of different,
particle-size-dependent reversal mechanism were published by
Wernsdorfer et al. \cite{wernsdorfer}.

As one important result, we found that in the limit of large particle
size the region where a thermally activated coherent rotation occurs
vanishes.  This means that for large particles the rotation is always
either nonthermal or driven by nucleation, depending only on the ratio
of the driving field to the anisotropy. Second, in the nucleation
regime the energy barrier is reduced, since here -- in contrast to the
N\'{e}el-Brown theory -- for vanishing magnetic field the energy
barrier is proportional to the cross-sectional area of the particle
rather than its volume.

We confirmed the result above by simulations. We also tried to
determine the relevant energy barriers numerically. For the case of
the Heisenberg model we could hardly reach the low-temperatures limit
which one needs in order to see the simplest, lowest-energy reversal
mechanism. However, for an Ising model, i.~e.~, in the limit of
infinite anisotropy we could establish our formulas for the energy
barrier of a single-droplet nucleation process.

{\bf Acknowledgments:} We thank K.~D.~Usadel for helpful suggestions
and critically reading of the manuscript. The work was supported by the
Deutsche Forschungsgemeinschaft through Sonderforschungsbereich 166
and through the Graduiertenkolleg "Heterogene Systeme".


\begin{references}
\bibitem{chantrell} R.~W.~Chantrell and K.~O'Grady, in  \emph{Applied
  Magnetism}, edited by R.~Gerber, C.~D.~Wright, and G.~Asti (Kluwer
  Academic Publishers, Dordrecht, 1994), p. 113
\bibitem{salling} C.~Salling, R.~O'Barr, S.~Schultz, I.~McFadyen, and
  M.~Ozaki. J.~Appl.~Phys. {\bf 75}, 7986 (1994)
\bibitem{lederman} M.~Lederman, S.~Schultz, and M.~Ozaki, Phys.~Rev.~Lett. {\bf 73}, 1986 (1994)
\bibitem{wernsdorfer} W.~Wernsdorfer, K.~Hasselbach, D.~Mailly,
  B.~Barbara, A.~Benoit, L.~Thomas, and G.~Suran,
  J.~Mag.~Mag.~Mat. {\bf 140}, 389 (1995)
\bibitem{wernsdorfer2} W.~Wernsdorfer, E.~Bonet Orozco, K.~Hasselbach,
  A.~Benoit, B.~Barbara, N.~Demoncy, A.~Loiseau, H.~Pascard, and D.~Mailly,
  Phys.~Rev.~Lett. {\bf 78}, 1791 (1997)
\bibitem{stauffer} D.~Stauffer, Int.~J.~Mod.~Phys. C {\bf 3}, No.5,
  1059 (1992)
\bibitem{tomita} H.~Tomita and S.~Miyashita, Phys.~Rev.~B {\bf 46},
  8886 (1992).
\bibitem{rikvold} P.~A.~Rikvold and B.~M.~Gorman, in  \emph{Annual Reviews of  Computational Physics I}, edited by D.~Staufer (World Scientific, Singapore, 1994), p. 149
\bibitem{acharyya} M.~Acharyya and B.~Chakrabarti, Phys.~Rev.~B {\bf
  52}, 6550 (1995)
\bibitem{garcia} D.~Garc\'{\i}a-Pablos, P.~Garc\'{\i}a-Mochales,
  N.~Garc\'{\i}a, and P.~A.~Serena, J.~Appl.~Phys. {\bf79}, 6019
  (1996)
\bibitem{richards} H.~L.~Richards, M.~Kolesik, P.~A.~Lindg{\aa}rd,
  P.~A.~Rikvold, and M.~A.~Novotny, Phys.~Rev.~B {\bf 55}, 11521 (1997)
\bibitem{neel} L.~N\'{e}el, Ann.~Geophys. {\bf 5}, 99 (1949)
\bibitem{brown} W.~F.~Brown, Phys.~Rev. {\bf 130}, 1677 (1963)
\bibitem{nowak} U.~Nowak, J.~Heimel, T.~Kleinefeld, and D.~Weller,
  Phys.~Rev.~B {\bf 56}, 8143 (1997)
\bibitem{aharony} A.~Aharoni, J.~Appl.~Phys. {\bf 63}, 5879 (1988)
\bibitem{becker} R.~Becker and W.~D\"{o}ring, Ann.~Phys. (Leipzig)
  {\bf 24}, 719 (1935)
\bibitem{braun} H.~B.~Braun, Phys.~Rev.~Lett. {\bf 71}, 3557 (1993)
\bibitem{binder} K.~Binder and H.~M\"{u}ller-Krumbhaar, Phys.~Rev.~B {\bf
  9}, 2328 (1974)
\bibitem{ray} T.~S.~Ray and J.~-S.~Wang, Physika A {\bf 167}, 580 (1990)
\bibitem{sides}  H.~L.~Richards, S.~W,~Sides, M.~A.~Novotny, and
  P.~A.~Rikvold, J.~Mag.~Mag.~Mat. {\bf 150}, 37 (1995)
\bibitem{heermann} K.~Binder and D.~W.~Heermann, \emph{Monte Carlo
  Simulation in Statistical Physics} (Springer-Verlag, Berlin, 1988), p. 21
\bibitem{reif} F.~Reif, \emph{Fundamentals of statistical and thermal
  physics} (McGraw-Hill Book Company, New York 1965), p. 548
\bibitem{ciftan} R.~G.~Brown and M.~Ciftan, Phys.~Rev.~Lett. {\bf 76},
  1352 (1996)

\end{references}
\end{document}